\documentclass[10pt, conference, compsocconf]{./IEEEtran}

\usepackage{booktabs}
\usepackage{graphicx}
\usepackage{amsmath}
\usepackage{amssymb}
\usepackage{color}
\usepackage{ifpdf}
\usepackage{float}
\usepackage[utf8]{inputenc}
\usepackage{multirow}
\usepackage{rotating}
\usepackage{subfigure}
\usepackage{array}

\usepackage{moresize}
\usepackage{url}
\usepackage{booktabs}
\usepackage{listings}   
\usepackage{paralist}    
\usepackage{wrapfig}    
\usepackage{multirow}
\usepackage{ifpdf}
\usepackage{xspace}
\usepackage{keyval}  
\usepackage{color}
\usepackage{cite}
\usepackage{flushend}

\definecolor{listinggray}{gray}{0.95}
\definecolor{darkgray}{gray}{0.7}
\definecolor{commentgreen}{rgb}{0, 0.4, 0}
\definecolor{darkblue}{rgb}{0, 0, 0.4}
\definecolor{middleblue}{rgb}{0, 0, 0.7}
\definecolor{darkred}{rgb}{0.4, 0, 0}
\definecolor{brown}{rgb}{0.5, 0.5, 0}
\definecolor{dkgreen}{rgb}{0,0.5,0}
\definecolor{orange}{rgb}{1,.5,0}
\definecolor{dandelion}{cmyk}{0,0.29,0.84,0}

\usepackage[normalem]{ulem}
\makeatletter
\def\cyanuwave{\bgroup \markoverwith{\lower3.5\p@\hbox{\sixly \textcolor{cyan}{\char58}}}\ULon}
\def\reduwave{\bgroup \markoverwith{\lower3.5\p@\hbox{\sixly \textcolor{red}{\char58}}}\ULon}
\def\blueuwave{\bgroup \markoverwith{\lower3.5\p@\hbox{\sixly \textcolor{blue}{\char58}}}\ULon}
\font\sixly=lasy6 
\makeatother

\newif\ifdraft
\ifdraft
 \newcommand{\N}[1]{\textbf{NOTE: #1}\xspace}
 \newcommand{\jhanote}[1]{ {\textcolor{cyan} { ***SJ: #1 }}}
 \newcommand{\katznote}[1]{ {\textcolor{blue} { ***DSK: #1 }}}
 \newcommand{\mtnote}[1]{ {\textcolor{orange} { ***MT: #1 }}}
 \newcommand{\jonnote}[1]{ {\textcolor{dkgreen} { ***JW: #1 }}}
 
 \newcommand{\note}[1]{ {\textcolor{red} { *** #1 }}}
\else
 \newcommand{\N}[1]{}
 \newcommand{\jhanote}[1]{}
 \newcommand{\katznote}[1]{}
 \newcommand{\mtnote}[1]{}
 \newcommand{\jonnote}[1]{}
 \newcommand{\note}[1]{}
\fi

\newif\ifdraft
\ifdraft
\newcommand{\terminology}[1]{ {\textcolor{red} {(Terminology used: \textbf{#1}) }}}

\newcommand{\alnote}[1]{ {\textcolor{blue} { ***andreL: #1 }}}
\newcommand{\amnote}[1]{ {\textcolor{blue} { ***andreM: #1 }}}
\newcommand{\smnote}[1]{ {\textcolor{brown} { ***sharath: #1 }}}
\newcommand{\pmnote}[1]{ {\textcolor{brown} { ***Pradeep: #1 }}}
\newcommand{\msnote}[1]{ {\textcolor{cyan} { ***mark: #1 }}}
\newcommand{\mrnote}[1]{ {\textcolor{purple} { ***melissa: #1 }}}
\else
\newcommand{\onote}[1]{}
\newcommand{\terminology}[1]{}

\newcommand{\alnote}[1]{}
\newcommand{\amnote}[1]{}
\newcommand{\athotanote}[1]{}
\newcommand{\smnote}[1]{}
\newcommand{\pmnote}[1]{}
\newcommand{\msnote}[1]{}
\newcommand{\mrnote}[1]{}
\newcommand{\aznote}[1]{}
\fi

\newcommand{\up}{\vspace*{-1em}}

\newcommand{\B}[1]{\textbf{#1}\xspace}

\newcommand{\mr}[1]{\multirow{2}{*}{#1}}%

\lstdefinestyle{myListing}{
  frame=single,   
  backgroundcolor=\color{listinggray},  
  language=C,       
  basicstyle=\ttfamily \footnotesize,
  breakautoindent=true,
  breaklines=true
  tabsize=2,
  captionpos=b,  
  aboveskip=0em,
  belowskip=-2em,
}      

\lstdefinestyle{myPythonListing}{
  frame=single,   
  backgroundcolor=\color{listinggray},  
  language=Python,       
  basicstyle=\ttfamily \scriptsize,
  breakautoindent=true,
  breaklines=true
  tabsize=2,
  captionpos=b,  
}

\ifpdf
\DeclareGraphicsExtensions{.pdf, .jpg, .tif}
\else
\DeclareGraphicsExtensions{.ps,  .eps, .jpg}
\fi

\begin{document}
\title{Integrating Abstractions to Enhance the Execution of Distributed
Applications}

\author{
  \IEEEauthorblockN{
    Matteo Turilli\IEEEauthorrefmark{1},
    Feng Liu\IEEEauthorrefmark{2},
    Zhao Zhang\IEEEauthorrefmark{3},
    Andre Merzky\IEEEauthorrefmark{1},\\
    Michael Wilde\IEEEauthorrefmark{4},
    Jon Weissman\IEEEauthorrefmark{2},
    Daniel S. Katz\IEEEauthorrefmark{4},
    Shantenu Jha\IEEEauthorrefmark{1}\IEEEauthorrefmark{5}
  }
  \IEEEauthorblockA{
    \IEEEauthorrefmark{1}RADICAL Laboratory, Electric and Computer Engineering, Rutgers University, New Brunswick, NJ, USA
  }
  \IEEEauthorblockA{
    \IEEEauthorrefmark{2}Computer Science and Engineering Department, University of Minnesota, Minneapolis, MN, USA
  }
  \IEEEauthorblockA{
    \IEEEauthorrefmark{3}AMPLab, University of California Berkeley, CA, USA
  }
  \IEEEauthorblockA{
    \IEEEauthorrefmark{4}Computation Institute, University of Chicago \& Argonne National Laboratory, Chicago, IL, USA
  }
 \IEEEauthorblockA{
    \IEEEauthorrefmark{5}Corresponding Author}
}

\maketitle

\begin{abstract}

  One of the factors that limits the scale, performance, and sophistication of
  distributed applications is the difficulty of concurrently executing them on
  multiple distributed computing resources. In part, this is due to a poor
  understanding of the general properties and performance of the coupling
  between applications and dynamic resources. This paper addresses this issue
  by integrating abstractions representing distributed applications, resources,
  and execution processes into a pilot-based middleware. The middleware
  provides a platform that can specify distributed applications, execute them
  on multiple resource and for different configurations, and is instrumented to
  support investigative analysis. We analyzed the execution of distributed
  applications using experiments that measure the benefits of using multiple
  resources, the late-binding of scheduling decisions, and the use of backfill
  scheduling.

\end{abstract}


\IEEEpeerreviewmaketitle

\section{Introduction}
\label{sec:intro}

Large-scale science projects~\cite{lsst-url,ska-url,lhc-url} as well as the
long-tail of science~\cite{hey2009fourth} rely on distributed computing
resources. While progress has been made in the use of individual resources, a
major challenge for high-performance and distributed computing
(HPDC)~\cite{hpdc_papers_url} is developing applications that can execute
concurrently on multiple resources~\cite{katz2011cyberinfrastructure}.

Resources are heterogeneous in their architectures and interfaces, are
optimized for specific applications, and enforce tailored usage and fair
policies. In conjunction with temporal variation of demand, this introduces
resource dynamism, e.g., the time varying availability, queue time, load,
storage space, and network latency.  Executing an application on multiple
heterogeneous and dynamic resources is difficult due to the complexity of
choosing resources and distributing the application's tasks over them.


Executing distributed applications on multiple dynamic resources requires
loosely coupling applications to a predetermined set of resources. This enables
applications to respond to changes in the availability of resources without
breaking predetermined application-resource assignments. Our work investigates
how to integrate information from both applications and resources to make
coupling decisions. In this paper, we study the static coupling between a
description of applications and information about dynamic resources.

Coupling applications to multiple dynamic resources requires advances at both
the conceptual and implementation level. We contribute by devising dedicated
abstractions, implementing them in middleware, and an experimental evaluation
to show the benefits of our methodology. The abstractions represent:
characterization of application tasks (skeletons); resource capacity and
capabilities (bundles); resource management and task scheduling (pilots); and
application execution (execution strategies).  Middleware that includes
implementations of these abstractions enables the execution of distributed
applications across multiple dynamic resources, and the experimental analysis
of the execution process and its performance.

We performed experiments over a year using up to five concurrent resources
belonging to XSEDE and NERSC, involving more than 20,000 runs of different
distributed applications for a total of 10 million tasks executed.  These
experiments measured the performance of alternative couplings between
distributed applications and multiple resources. Each coupling is described by
an execution strategy, i.e., the set of decisions made to execute an
application.  Each strategy is characterized by a time to distribute and
execute the application tasks across the resource. We found that execution
strategies based on late-binding and backfilling of tasks on at least three
resources offered the best performance. We observed this to be independent of
the combination of resources used, the number of application's tasks, and the
distribution of the tasks' duration.

This work advances traditional reasoning about distributed execution by
performing both hypothesis-driven and semi-empirical investigations. The
abstractions advance analytical understanding of coupling distributed
applications and resources; the middleware implementation advances the
state-of-the-art of executing specific distributed applications on multiple
dynamic resources; and experiments improve understanding of the execution
process and its performance.

In Section 2, we provide a brief summary of related work. Section 3 discusses
the abstractions we defined alongside the architecture and resultant
capabilities of implementing the abstractions into an integrated middleware.
Section 4 discusses the experimental methodology and design, the results of the
experiments, and their analysis. Section 5 reviews the implications and
challenges of the presented work alongside its implications for the future of
HPDC.

\section{Related Work}
\label{sec:rwork}

Coupling of distributed applications to multiple resources is a well-known
research problem. For example, the I-Way~\cite{foster1996software},
Legion~\cite{grimshaw1997legion}, Globus~\cite{foster2001anatomy}, and
HTCondor~\cite{thain2005distributed} frameworks integrated existing tools to
run distributed applications on multiple resources. However, their
successes~\cite{allen2001supporting,gardner2006parallel} also exposed their
limitations. Deploying mandatory middleware on all resources and the complexity
of porting applications limited adoption and
usability~\cite{Jha2013distributed}. The abstractions and middleware presented
in this paper avoid these limitations by: (1) implementing interoperability
among resource middleware; (2) assuming multiple resources with dynamic
availability and diverse capabilities; and (3) understanding and measuring
performance of alternative coupling between distributed applications and those
resources.

The Execution Strategy, Skeleton, and Bundle abstractions build upon related
work. Bokhari~\cite{bokhari1981shortest} and
Fernandez-Baca~\cite{fernandez1989allocating} theoretically proved the NP
completeness of matching/scheduling components on distributed systems. Chen and
Deelman~\cite{chen2012partitioning}, and Malawski et
al.~\cite{malawski2015algorithms} present execution strategies modeled as
heuristics based on empirical experimentation.  Workflow systems like Kepler,
Swift and Taverna~\cite{taylor2006} implement execution management but in the
form of a single point solution, specifically tailored to their execution
models. Work from Foster and Stevens~\cite{foster1990parallel} and  Meyer et
al.~\cite{meyerwgl} resemble some of the features of skeletons but are limited
only to parallel applications or directed acyclic graphs (DAGs). Bundles
leverages some of the work done in information
collection~\cite{cardosa2008resource}, resource
discovery~\cite{liu2003constraint,oppenheimer2005design}, and resource
characterization as it relates to queue time prediction~\cite{nurmi2008qbets}
and its difficulties~\cite{tsafrir2007backfilling}.


\section{Abstractions}
\label{sec:abstractions}

Executing a distributed application requires information about the application
requirements, and resource availability and capabilities. This information is
used to choose a suitable set of resources on which to run the application
executable(s) and a suitable scheduling of the application tasks. When
considering multiple resources and distributed applications, bringing together
application- and resource-level information requires specific abstractions that
have to uniformly and consistently describe the core properties of distributed
applications, those of computing resources, and those of the execution of the
former on the latter.


\subsection{Application Abstractions}

Most distributed applications that we have observed are of two types. The first
is Many-Task Computing (MTC)~\cite{raicu2008many}, where applications are
composed of tasks, which themselves are executables. The tasks can be thought
of as having a simple structure from the outside: they read files, compute, and
write files, though they can be quite complex internally. They can also be
either sequential or parallel.  These MTC applications fall into a small number
of types, specifically bag-of-task, (iterative) map-reduce, and (iterative)
multistage workflow.  Interaction between tasks is usually of the form of files
produced by one task and consumed by another.  The tasks can be distributed
across resources, and there is a framework that is responsible for launching
the tasks and moving the files as needed to allow the work to be done.

The second type is applications composed of distributed elements that interact
in a more complex manner, such as by exchanging messages while running,
possibly as services.  The elements of these applications can have persistent
state, while the MTC tasks do not; their outputs are based solely on their
inputs, and they are basically idempotent.

The majority of distributed science and engineering applications are MTC, while
in business, there is much more of a mix of the types. Because we are concerned
with science and engineering application, we currently focus on MTC
applications. We generalize bag-of-task, (iterative) map-reduce, and
(iterative) multistage workflow applications into (iterative) multistage
workflow applications, since bag-of-task applications are basically
single-stage applications and map-reduce applications are basically two-stage
applications.

We abstract these applications because the real applications can be difficult
to obtain and to build, the real input data sets may be difficult to obtain,
and the real applications may be difficult to arbitrarily scale. Abstract
applications also can be easily shared and are reproducible. In order to
abstract these applications, we use a top-down approach: an application is
composed of a number of stages (which can be iterated in groups), and each
stage has a number of tasks. An application is described by specifying the
number of stages and the number of tasks, input and output file and task
mapping, task length, and file size inside each stage. Task lengths and file
sizes can be statistical distributions or polynomial functions of other
parameters. For example, input file size can be a normal distribution, task
length can be a linear function of input file size, and output size can be a
binomial function of task runtime.

We call this type of abstract application a ``Skeleton
Application''~\cite{zhang2014using, skeleton-FGCS}, and have built an open
source tool (Application Skeleton~\cite{skeleton-code-v1.2}) that can create
skeletons. Our tool is implemented as a parser that reads in a configuration
file that specifies a skeleton application, and produces three groups of
outputs: (1) \textbf{Preparation Scripts}: run to produce the input/output
directories and input files for the skeleton application. (2)
\textbf{Executables}: the actual tasks of each application stage. (3)
\textbf{Skeleton Application}: implemented as: (a) shell commands that can be
executed in sequential order on a single machine, (b) a Pegasus
DAG~\cite{deelman2005pegasus} or (c) a Swift script~\cite{wilde2011swift} that
can be executed on a local machine or in a distributed environment, or (d) a
JSON structure that must be used by a middleware that is designed to read it.

The application skeleton tool itself can be called from the command line or
through an API. Our work here uses the skeleton API to call the skeleton code.
To execute a skeleton application, the preparation scripts are run to create
the initial input data files, then the skeleton application itself is run. The
task executables produced by the skeleton tool copy the input files from the
file system to RAM, sleep for some amount of time (specified as the runtime),
and copy the output files from RAM to the file system.

We have previously tested the performance accuracy of the skeleton
applications~\cite{zhang2014using}. We profiled three representative
distributed applications---Montage~\cite{jacob2009montage},
BLAST~\cite{mathog2003parallel},
CyberShake-postprocessing~\cite{maechling2007scec}---then derived appropriate
skeleton parameters. We showed that the application skeleton tool produced
skeleton applications that correctly captured important distributed properties
of real applications but were much simpler to define and use. Performance
difference between the real applications vs the skeleton applications were
-1.3\%, 1.5\%, and 2.4\%. Fourteen out of fifteen application stages had
differences of under 3\%, ten had differences under 2\%, and four had
differences of under 1\%.

\subsection{Resource Abstractions}

Very few scientific applications run on dedicated resources owned by the end
user. As a consequence, most of the users of these applications have to share
resources that have dynamic availability and diverse capabilities. This
introduces complexity for both the application developer and end user. We
mitigate this complexity with a resource abstraction.

Resource allocation for applications can be either static or dynamic.
Allocation is static when users select resources based on knowledge of
capacity, performance, policy, and cost. Often, this decision is made on an ad
hoc basis. Allocation is dynamic when users, or code running on their behalf,
monitor resource status and adjust resources as needed. Both static and dynamic
resource allocation require resource characterization but despite its
importance, we observed that we lack systematic approaches to such a
characterization.

We developed an abstraction to bridge applications and diverse resources via
uniform resource characterizations. In this way, we facilitate efficient
resource selection by distributed applications. Our resource abstraction is
called ``Bundle'' to connote the characterization of a collection of resources.


Our implementation of the Bundle abstraction is the {\em resource bundle},
which may be thought of as representing some portion of system resources. A
resource bundle may contain an arbitrary number of resource categories (e.g.,
compute, storage) but it does not ``own'' the resources. In this way, a
resource may be shared across multiple bundles and users can be provided with a
convenient handle for performing aggregated operations such as querying and
monitoring. Resource bundles are used to enable applications to make more
effective resource allocation decisions.

A resource bundle has two components: {\em resource representation} and {\em
resource interface}. The resource representation characterizes heterogeneous
resources with a large degree of uniformity, thus hiding complexity. Currently,
the resource bundle models resources across three basic categories: compute,
network, and storage. Resource measures that are meaningful across multiple
platforms are identified in each category. For example, the property ``setup
time'' of a compute resource means queue wait time on a HPC cluster or virtual
machine startup latency on a cloud~\cite{simmhan2010comparison}.

The resource interface exposes information about resources availability and
capabilities via an API. Two query modes are supported: on-demand and
predictive. The on-demand mode offers real-time measurements while the
predictive mode offers forecasts based on historical measurements of resource
utilization instead of queue waiting time, which is extremely hard to predict
accurately~\cite{tsafrir2007backfilling,downey1997predicting,nurmi2008qbets}.

The resource interface exposes three types of interface: querying, monitoring,
and discovering. The query interface uses end-to-end measurements to organize
resource information. For example, the query interface can be used to inquire
how long it would take to transfer a file from one location to a resource and
vice versa. Although file transfer times are difficult to
estimate~\cite{vazhkudai2001predicting}, proper tools~\cite{kosar2004stork} are
capable of providing estimates within an order of magnitude, which are still
useful.

The monitoring interface can be used to inquire about resource state and to
chose system events for which to receive notification. For example, performance
variation within a cluster can be monitored so that when the average
performance has dropped below a certain threshold for a certain period,
subscribers of such an event will be notified. This may trigger subsequent
scheduling decisions such as adding more resources to the application.

The discovery interface, which is future work, will let the user request
resources based on abstract requirements so that a tailored bundle can be
created. A language for specifying resource requirements is being developed.
This concept has been shown to be successful for storage aggregates in the
Tiera project~\cite{raghavan2014tiera}, where resource capacities and resource
policies are specified in a compact notation.

\subsection{Dynamic Resource Abstractions}

Pilots generalize the common concept of a resource placeholder. A pilot is
submitted to the scheduler of a resource, and once active, accepts and executes
tasks directly submitted to it. In this way, the tasks are executed within the
time and space boundaries set by the resource's scheduler for the pilot,
trading the scheduler overhead for each task with an overhead for a single
pilot.

Pilots have proven very successful at supporting distributed applications,
especially those with large scale, single/low cores tasks. For example, the
ATLAS project~\cite{atlas2008atlas} processes 5 million jobs every
week~\cite{maeno2014evolution} with the Production and Distributed Analysis
(PanDA) system and uses pilots to execute single-task jobs. The Open Science
Grid (OSG)~\cite{pordes2007open} deploys HTCondor and a pilot system named
``Glidein'' to make available 700 million CPU hours a year for applications
requiring high-throughput of single-core tasks~\cite{katz2012survey}. Pilots
are also used by several user-facing tools to execute workflows. Swift
implements pilots in a subsystem named ``Coasters''~\cite{hategan2011coasters},
FireWorks employs ``Rockets''~\cite{jain2015fireworks}, and Pegasus uses
Glidein via providers like ``Corral''~\cite{deelman2005pegasus}.

Most existing pilot systems are part of resource-specific middleware or of a
vertical application framework; they have also not been instrumented so as to
provide accurate information about their internal state and operations. To
avoid these limitations, we utilize and extend
RADICAL-Pilot~\cite{review_radicalpilot_2015}, a pilot system
that does not need to be deployed in the resource middleware and exposes a
well-defined programming interface to user-facing tools. RADICAL-Pilot uses
RADICAL-SAGA~\cite{saga-x} (the reference implementation of the SAGA OGF
standard~\cite{goodale2008simple}) to submit pilots and execute tasks on
multiple resources. Timers and introspection tools record each state transition
and the state properties of each RADICAL-Pilot component. These capabilities
are needed to tailor distributed application execution to diverse use cases,
but to the best of our knowledge, they are missing in other pilot systems.
RADICAL-Pilot also adheres to recent advances in elucidating the pilot
paradigm~\cite{piloterview-2015}.

\subsection{Execution Abstractions}

Coupling applications and resources is a matching process that depends on
information and a set of decisions. Information about the application
requirements and both the resources availability and capabilities is collected.
This information is then integrated and used to take decisions about the amount
and type of resources that are needed to satisfy the application requirements,
for how long these resources need to be available, how data should be managed,
and how the application should be executed on the chosen resources.

We use ``Execution Strategy'' to refer to all the decisions taken when
executing a given application on one or more resources. We use this set of
decisions to describe the process of coupling applications to resources.
Execution Strategy is the abstraction while the set of choices made for each of
its decision is one of its realizations. We call these realizations ``execution
strategies'' or simply ``strategies''.


We use the Execution Strategy abstraction to make explicit the decisions that,
traditionally, remain implicit in the coupling of applications and resources.
In the presence of multiple dynamic resources, individual user knowledge and
experience or best practices are not sufficient to identify alternative ways to
couple an application to resources, and to understand their performance trade
offs. Once the decisions are made explicit, they can be integrated into a model
and their effects can be measured empirically.

An Execution Strategy can be thought of as a tree, where each decision is a
vertex and each edge is a dependence relation among decisions. In the simplest
case, there is only one choice (i.e., value) for each decision that enables the
execution of the application. When considering multiple resources, the number
of choices, the information needed to make those choices, and the complexity of
the decision process all increase. For example, resources can be used
concurrently; the distribution of tasks among resources can be uneven;
multi-level scheduling may be used for late binding, or alternative scheduling
algorithms may be available. The dependency among decisions defines their
sequence. For example, the decision to schedule tasks on each resource may
depend on the number of resources that has been chosen.

Once enacted, each execution strategy executes an application, but different
strategies lead to better or worse performance, as measured using diverse
metrics. For example, execution strategies may differ in terms of
time-to-completion (TTC), throughput, energy consumption, affinity to specific
resources, or economic considerations. Each metric's relevance depends on the
user's and application's qualitative requirements. TTC is one of the most
relevant metrics; we investigated distributed application execution by
experimenting whether, how, and why alternative execution strategies lead to
different TTC for the same application.

Execution Strategies are realized in a software module called ``Execution
Manager''. This module derives and enacts an execution strategy in five steps:
(1) information is gathered about an application via the skeleton API and about
resources via the bundle API; (2) application requirements and resources
availability and capabilities are determined; (3) a set of suitable resources
is chosen to satisfy the application requirements; (4) a set of suitable pilots
is described and then instantiated on the chosen resources; and (5) the
application is executed on the instantiated pilots.

The choices made in steps 3 and 4 depend on whether an optimization metric is
given to the Execution Manager. For example, given a distributed application
composed of independent tasks and the TTC metric, the Execution Manager selects
a set of resources to achieve maximal execution concurrency and minimize the
execution time. Similar decisions are made depending on the amount of data that
needs to be staged, the bandwidth available between resources, the set of data
dependences, etc. Note that this type of optimization uses semi-empirical
heuristics. For example, algorithmic considerations and empirical evidence
about pilots and resources behavior need to be known to decide whether to
execute: (1) a bag of 2048 single-core, loosely-coupled tasks, with a varying
number of cores, duration, and input/output data on a single large pilot; or
(2) the same bag of tasks on three smaller pilots instantiated on different
resources.

\subsection{Integration}
\label{sec:integration}

\begin{figure}[!t]
    \centering
    \includegraphics[width=0.40\textwidth]{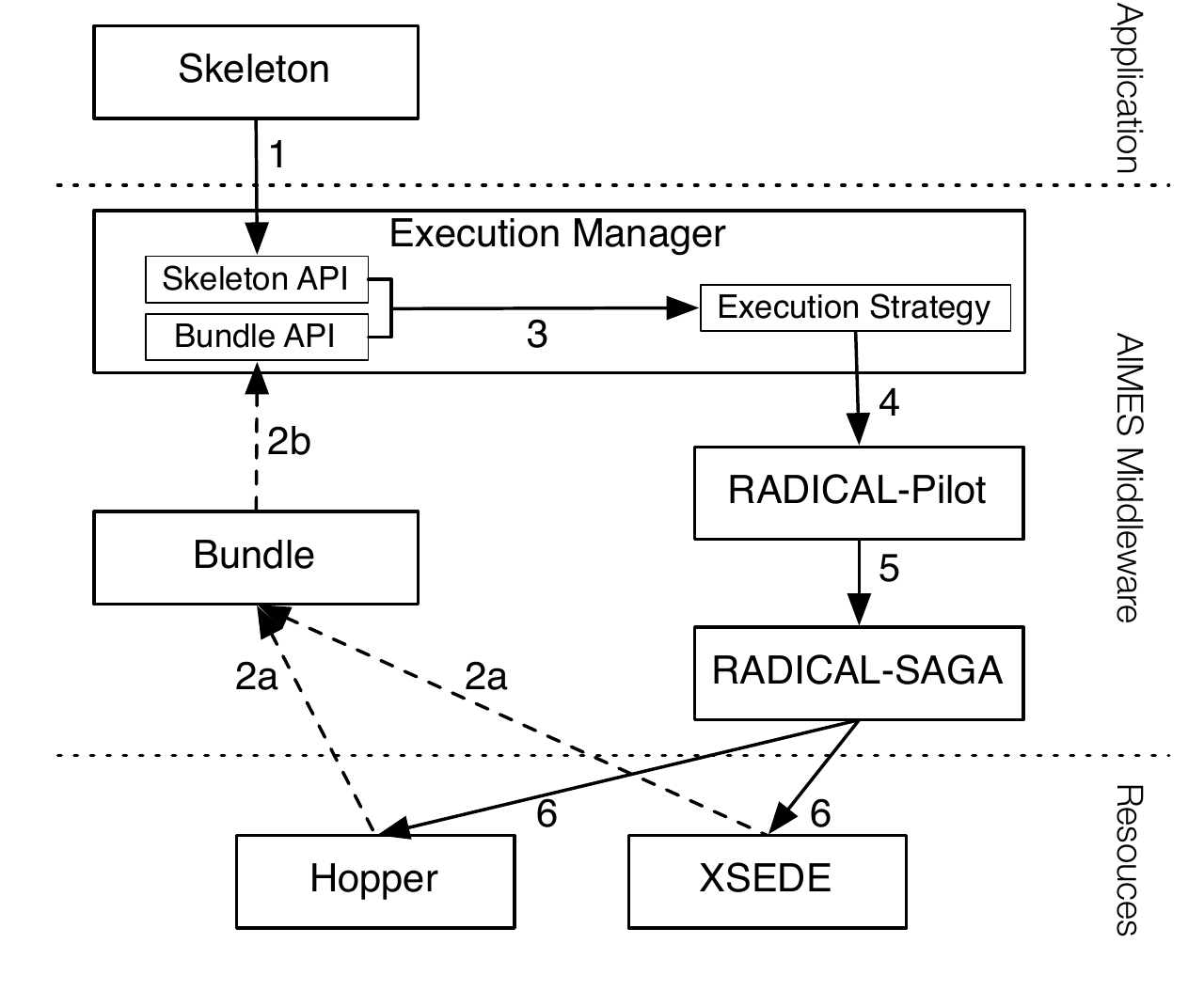}
    \caption{High-level architecture of the AIMES middleware. (1) provides
    distributed application description; (2) provides resource availability and
    capabilities; (3) derives execution strategy; (4-6) enacts execution
    strategy.
    \label{fig:aimes_architecture}
    \up}
\end{figure}

We implemented the four abstractions---Skeleton Application, Bundle, Pilot, and
Execution Strategy---as Python modules, then integrated them into the AIMES
(Abstractions and Integrated Middleware for Extreme-Scales) middleware (see
Figure~\ref{fig:aimes_architecture}). This middleware offers two distinguishing
features: self-containment, meaning no components need to be deployed into the
resources, and self-introspection, meaning that its state model is explicit and
instrumented to produce complete traces of an application execution.

The AIMES middleware realizes the coupling of a distributed application to
multiple dynamic resources. The coupling is described via an execution strategy
and then enacted via the pilot and interoperability modules. In this way, the
AIMES middleware integrates the implementations of the presented abstractions
but also the information about application and resources required by their
coupling. Thanks to the self-introspection, the AIMES middleware can work as an
experimental laboratory: The enactment of the coupling process can be measured
in all its stages to understand the correlations among the choices of an
execution strategy and the performance of the execution process.

The AIMES middleware exposes capabilities at the application, resource, and
execution management level. We use the skeleton module to describe a
distributed application composed of many executable tasks by specifying their
number, duration, intercommunication requirements, data dependences, and
grouping into stages. We also use the bundle module to describe a set of
resources. Bundle information includes resource capacity, configuration, and
availability in terms of utilization, queue state, queue composition, and types
of jobs already scheduled for execution. We integrate the application and
resource information deriving an execution strategy and enacting it via
RADICAL-Pilot.

The AIMES middleware uses the skeleton API to acquire the application's
description (Figure~\ref{fig:aimes_architecture}, step 1). The same is done for
the resources via the bundle API (Figure~\ref{fig:aimes_architecture}, step
2a/b). The Execution Manager derives an execution strategy
(Figure~\ref{fig:aimes_architecture}, step 3) and then enacts it
(Figure~\ref{fig:aimes_architecture}, step 4, 5, 6). Pilots are described via
RADICAL-Pilot (Figure~\ref{fig:aimes_architecture}, step 4) and scheduled on
the chosen resources via RADICAL-SAGA (Figure~\ref{fig:aimes_architecture},
step 5). Once the pilots become active, tasks' input files are staged on the
resources of the active pilots and then tasks are scheduled and executed on
those pilots (Figure~\ref{fig:aimes_architecture}, step 6). Tasks are
automatically restarted in case of failure and, once executed, task output(s)
are staged back to the source where the AIMES middleware is being used.
Finally, all pilots are canceled when all tasks have executed so as not to
waste resources.


\section{Experiments}
\label{sec:exp}

We used the AIMES middleware to analyze alternative couplings of applications
and multiple dynamic resources, and to measure their performance. We described
the couplings by means of alternative execution strategies and observed the
relation between the choices of each execution strategy and the dynamic
behavior of the multiple resources. We acquired data over one year,
measuring experiment performance on four XSEDE and one NERSC resources.

\subsection{Methodology and Design}

We compare the performance of our execution strategies by measuring
applications TTC: the sum of a set of possibly overlapping time components.
Examples of time components are the time taken by each task to execute or the
time taken for a pilot waiting in a queue to be instantiated.  These components
can overlap as tasks can execute concurrently, and a pilot can be queued while
tasks execute on an active pilot. Differences in TTC depend on the duration and
overlap of its time components: the shorter the time component and/or the
higher their overlap, the lower the TTC.

We instrumented the AIMES middleware to record every TTC time component related
to middleware overhead, resource dynamism, task execution, and data staging. We
isolated the differences in TTC arising from the choices in an execution
strategy by running experiments with the same set of tasks and different
execution strategies. In this way, by comparing the time components of the TTC,
we gained insight into whether and how the execution strategies differ in the
duration of time components and/or their overlap.


\begin{table*}[!t]
\caption{Skeleton applications and execution strategies used for the
  experiments. Each application task runs on a single core. $T_x =$~estimated
  workflow execution time; $T_s =$ estimated total data staging time; $T_{rp}
  =$ AIMES middleware overhead.}
\label{table:strategies}
\centering
\begin{tabular}{@{}clllllll@{}}
\toprule
\B{Experiment} & \multicolumn{2}{c}{\B{Skeleton Application}} & \multicolumn{5}{c}{\B{Execution Strategy}} \\
\cmidrule(r){2-3}\cmidrule(r){4-8}
\B{ID} & \B{\#Tasks} & \B{Task Duration (min)} & \B{Binding} & \B{Scheduler} & \B{\#Pilots} & \B{Pilot Size} & \B{Pilot Walltime} \\
\midrule
\B{1} & $2^n, n=[3,11]$ & $15$ & \mr{Early} & \mr{Direct} & \mr{$1$} & \mr{$\#Tasks$} & \mr{$T_x + T_s + T_{rp}$} \\
\B{2} & $2^n, n=[3,11]$ & $1-30$ (trunc. Gaussian) &  & & & & \\
\midrule
\B{3} & $2^n, n=[3,11]$ & $15$ & \mr{Late} & \mr{Backfill} & \mr{$1-3$} & \mr{$\frac{\#Tasks}{\#Pilots}$} & \mr{${(T_x + T_s + T_{rp})}\cdot{\#Pilots}$} \\
\B{4} & $2^n, n=[3,11]$ & $1-30$  (trunc. Gaussian) & & & & & \\
\bottomrule
\end{tabular}
\end{table*}

We focused our experiments on the time components related to the resource
dynamism. Specifically, we measured and analyzed the advantage in performance
that can be obtained by using multiple pilots, late-binding, and back-filling
scheduling to increase execution overlapping while minimizing queuing time
duration.

We designed four experiments, each measuring the impact of an execution
strategy on 9 skeletons (see Table~\ref{table:strategies}). Each skeleton is a
distinct application that belongs to the same application class (bag-of-task)
but differs in size as measured by the number of tasks. Applications vary in
size between 8 and 2048 single-core tasks, and with task length of 15 minutes
or distributed following a truncated Gaussian (mean: 15 min.; stdev: 5 min.;
bounds: [1--30 min.]). The execution strategy decisions are: (1) early or late
binding of tasks to pilots; (2) the type of scheduler used to place tasks on
pilots; (3) the number of pilots; (4) their size; and (5) their walltime.

The size of the application and the durations of its tasks were selected to be
consistent with several scientific applications that could benefit from
distributed execution~\cite{bfe-jctc-2014}. Furthermore, the chosen durations
are also consistent with about 35\% of the jobs executed on XSEDE every year.
XDMoD~\cite{furlani2013using} shows that in 2014, more than 13 million jobs
were executed on XSEDE with durations between 30s and 30m, 36\% of the total
XSEDE workload. Between 2010-2013, 25\% to 55\% of the XSEDE workload was
within our experimental parameters.


Because skeletons have been shown to replicate the behavior of real-life
applications like Montage, Blast, and CyberShake with minimal
error~\cite{zhang2014using}, we used Skeleton workloads to enable better
control over the workload parameters and over the exploration of regions of
decision space.


Each experiment combined an execution strategy with one skeleton and its nine
applications. In total the four experiments consisted of 36 unique
applications. Each application was run many times depending on run-to-run
fluctuation. The execution order of the 36 applications was varied to avoid
correlation between measurements. The applications were also executed at
irregular intervals so as to avoid effects of short-term resource load
patterns. The order in which pilots were submitted to the resources was
randomized to account for differences in submission time across resources.

Table~\ref{table:strategies} shows that we selected a subset of all the
possible combinations between the given applications parameters and the choices
available for the given execution strategy. We discarded the following
combinations because they are redundant, uninformative, or ineffective: early
binding and multiple pilots; late binding and multiple pilots with enough cores
to execute all the tasks concurrently; early/late binding on pilots with the
same walltime; early/late binding with the same schedulers.

In early binding tasks are bound to the pilots before they become active. Given
more than one pilot, the TTC of execution strategies with early binding is
determined by the last pilot becoming active and all its tasks being executed.
In late binding the tasks are instead bound to pilots as soon as they become
active. In the worse case, the TTC of executions with late binding depends on
the time taken to execute all the tasks on the first available pilot. Given the
same tasks and the same pilots, the TTC of an execution strategy with early
binding is always equal or greater than one with late binding. They are equal
when multiple pilots become active at the same time; when pilots become active
at different times, TTC of an execution strategy with early binding is greater
than with late binding. We therefore perform experiments with early binding and
only one pilot.

Late binding with multiple pilots is a special case of early binding with a
single pilot when the pilots have enough cores to concurrently execute all the
tasks. With both bindings, all the tasks are executed on a single pilot as soon
as it becomes available. The remaining two pilots do not contribute to the TTC
of the application and are an overuse of resources. Accordingly, we use late
binding and pilots with fewer cores that those required to execute all the
given tasks. As a consequence, some of the application's tasks are executed
sequentially requiring more time compared to a fully concurrent execution. The
pilots we use for early and late binding have therefore different walltimes.

Experiment with early and late binding on a single pilot have the same TTC
because all the tasks are executed as soon as the pilot becomes active. As
such, we do not perform experiments with late binding and a single pilot. We
also use a single scheduler for early binding because using different
schedulers would measure the performance of the scheduler implementation,
something not directly related to the problem of coupling an application to
multiple resources of multiple DCI. Analogously, we do not perform experiments
to measure the performance of different schedulers for late binding.

\subsection{Results and Analysis}

\begin{figure}[]
    \centering
    \includegraphics[width=0.40\textwidth]{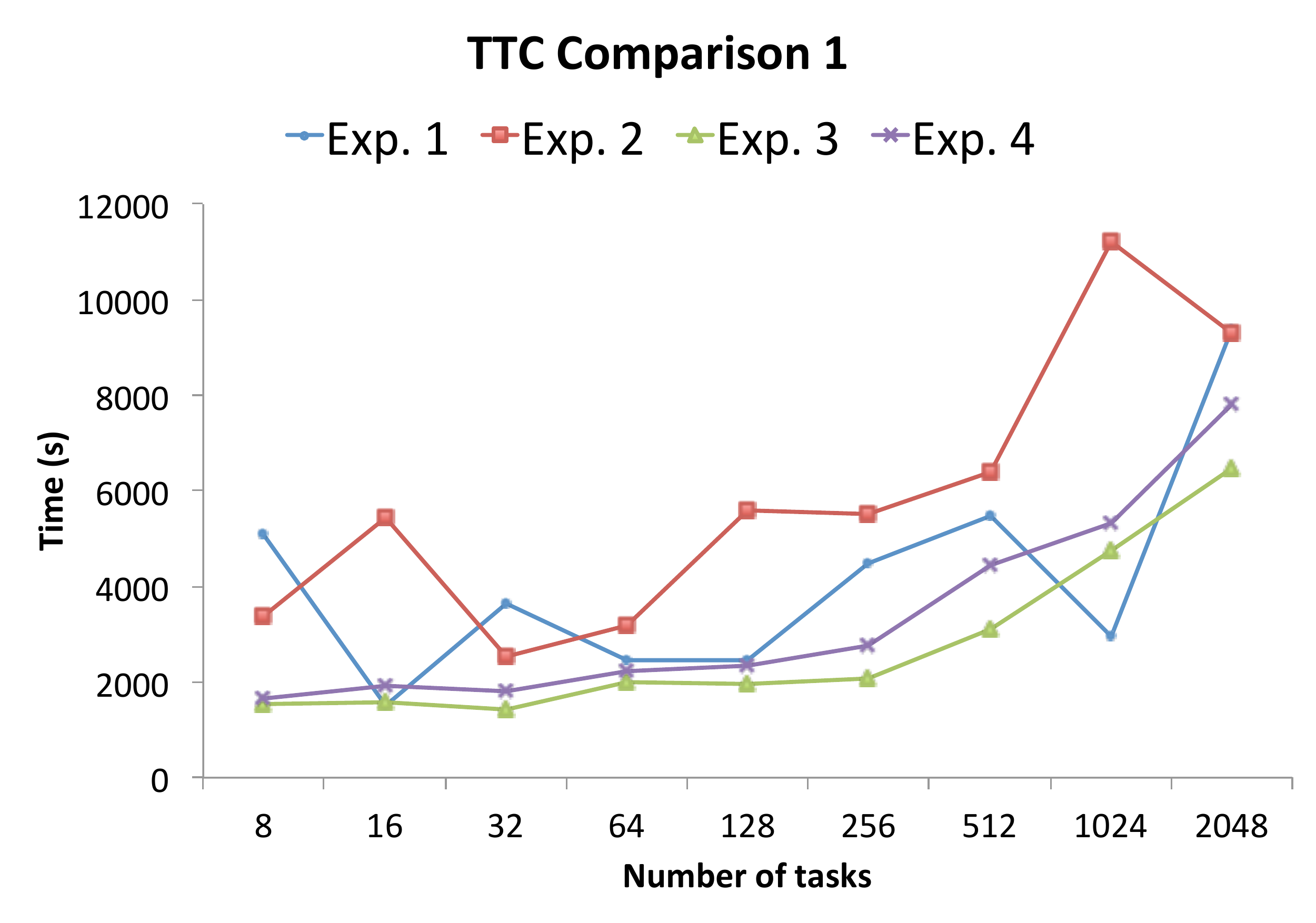}
    \caption{Comparison of TTC for experiments 1-4 shows large variations of
    the TTC in experiment 1 and 2 and smooth progression of TTC in experiment 3
    and 4.
    \label{fig:ttc_comparison}}
\end{figure}

We analyzed the experimental data comparing the performance of different
execution strategies for the given applications. We explained the observed
performance differences by isolating the components of the execution process
and measuring how their duration and overlapping contributed to the execution
TTC. Figure~\ref{fig:ttc_comparison} shows that for the strategies listed in
Table~\ref{table:strategies}, those using late binding and backfill scheduling
(Exp. 3 \& 4, green and purple lines) have shorter TTC, and therefore perform
better, than strategies using early binding and direct scheduling (Exp. 1 \& 2,
azure and red lines).

\begin{figure*}[t!]
    \centering
    \includegraphics[width=1\textwidth]{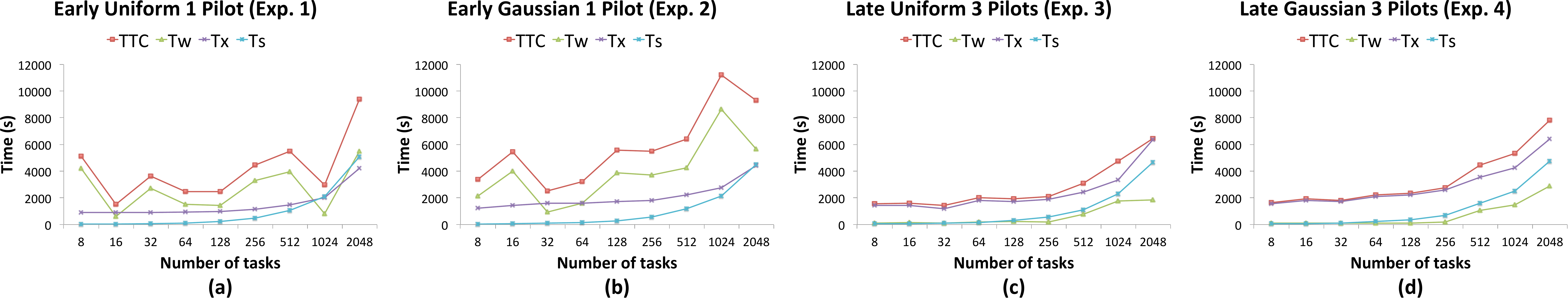}
    \caption{TTC and its time constituents presented for each experiment in
      Table~\ref{table:strategies} as a function of the distributed application
      size. $T_w$ = pilot setup and queuing time; $T_x$ = execution time; $T_s$
      = input/output files staging time. During execution ($T_w$), ($T_x$), and
      ($T_s$) overlap so TTC $< T_w + T_x + T_s$.
      \label{fig:ttc_composition}}
\end{figure*}

Figure~\ref{fig:ttc_composition} shows the three main components of TTC for
each experiment: (1) time setting up the execution including waiting for the
pilot(s) to become active on the target resource(s) ($T_w$, green); (2) time
executing all the application tasks on the available pilot(s) ($T_x$, purple);
and (3) time staging application data in and out ($T_s$, azure).

$T_s$ depends on the application specifications and has been restricted to a
small percentage of the overall TTC by experimental design. Larger amounts of
data could make $T_s$ dominant (Fig.~\ref{fig:ttc_composition}) and a set of
dedicated experiments would be required to investigate the differences among
strategies with decisions about, for example, compute/data affinity, amount of
network bandwidth available between the origin of the data and the target
resource(s), or the number and location of data replicas. This is the subject
of future work.

$T_x$ also depends on the application specifications. Our task durations are
either 15 minutes or a truncated Gaussian distribution between 1 and 30
minutes. However, $T_w$ depends mostly on the resource's queuing time.
This is determined by the resource load, the length of its queue, and
the policies regulating priorities among jobs and usage fairness among users.
As such, $T_w$ is outside user and middleware control.

Looking at how $T_s$, $T_x$, and $T_w$ contribute to the application TTC, we
note that in Figure~\ref{fig:ttc_composition}, $T_s$ is consistent across the
four execution strategies, contributing to the TTC proportionally to the number
of tasks executed. This is by design as every task requires a single input file
of 1~MB and produces a single output file of 2~KB. As a consequence, the more
tasks are executed, the more time is spent staging input and output files.

$T_x$ varies mostly over bindings, with the late binding strategies requiring
roughly 1/3 more time on average to execute tasks than those using early
binding. This is also by design as the late binding strategies use up to 3
pilots, each one with 1/3 of the cores used by the pilot of the early binding
strategies. As expected, the contribution of $T_x$ to the TTC is proportional
to the number of tasks executed but it increases with a steeper gradient above
256 tasks due to the overheads introduced by the AIMES middleware.

$T_w$ is the TTC component with the most variation among
experiments and the most contribution to TTC. The variation is more
pronounced for early binding (between 600 and 8,600 seconds) less for late
binding (between 99 and 2,800 seconds). In early binding, $T_w$ has large
variation vs. number of tasks, and between uniform and Gaussian distribution of
task durations. In late binding, $T_w$ has a smooth increase vs. number of
tasks and almost no variation between uniform and Gaussian distributions of
task durations.

Figure~\ref{fig:ttc_composition} shows also the dominance of $T_w$ contribution
to TTC. In the early binding strategies (Figures~\ref{fig:ttc_composition} (a)
and (b)), the red line (TTC) and the green line ($T_w$) have the same shape,
with the variations in TTC determined by equivalent variations in $T_w$. $T_w$
dominates also the TTC of the late binding strategies.
Figures~\ref{fig:ttc_composition} (c) and (d) show how $T_w$ accounts for
almost all the TTC for runs with up 256 tasks, and most of the TTC for the
remaining number of tasks.

\begin{figure}[!b]
    \centering
    \includegraphics[width=0.45\textwidth]{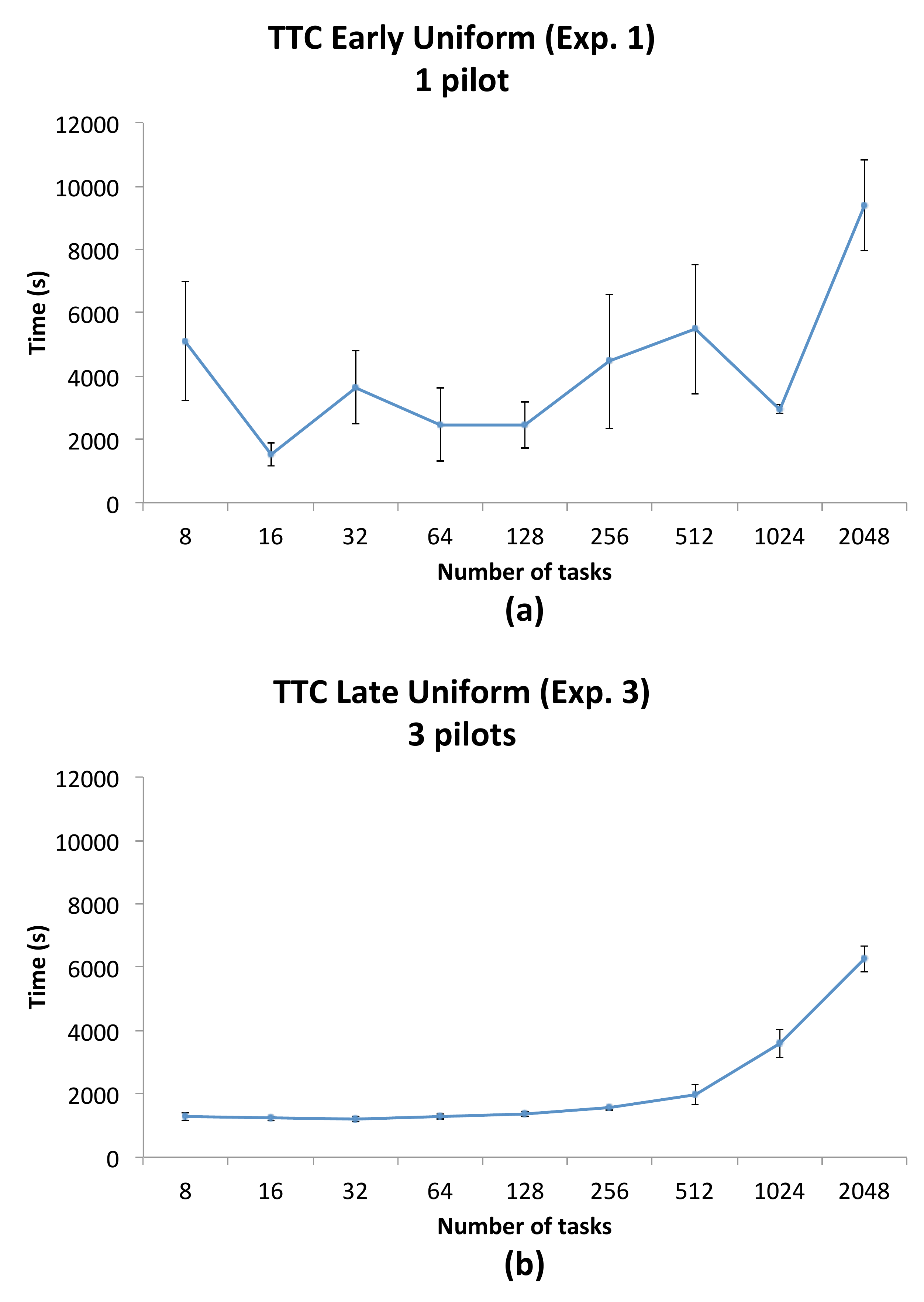}
    \caption{Figures (a) and (b) show the TTC for early and late binding.
    Differences in the size of the relative errors in (a) and (b) are
    consistent with the variance of $T_w$ observed in
    Figure~\ref{fig:ttc_composition}.
    \label{fig:ttc_comparison_vertical_TTC}}
\end{figure}

The difference in performance across strategies is therefore due to the
duration of $T_w$. On average, when using three pilots, the first pilot takes
less time to become active than when using a single pilot.
Figures~\ref{fig:ttc_comparison_vertical_TTC} (a) and (b) offer a possible
explanation of this behavior. The large error bars of
Figure~\ref{fig:ttc_comparison_vertical_TTC} (a) shows the variability of $T_w$
for the same job submitted multiple times to the same resource with early
binding. The variability between uniform and Gaussian distributions at 16 and
1024 tasks (Figure~\ref{fig:ttc_comparison}) is also a strong indication of how
widely TTC for early binding varies.
Figure~\ref{fig:ttc_comparison_vertical_TTC} (b) shows instead small error bars
across all task sizes. Due to the large variation in $T_w$ on each single
resource, when using multiple resources, it is more likely to find at least one
of the resources with a comparatively short $T_w$.

It is interesting that this large variability is already overcome by using
three resources (pilots) as shown in
Figure~\ref{fig:ttc_comparison_vertical_TTC} (b). Depending on availability,
the resources are chosen from a pool of five resources that are not only
dynamic as shown by the variability of $T_w$ but also heterogeneous in their
interfaces, size, utilization patterns, and possibly job priority and fair
usage policies. This hints to a generality of the observed behavior but future
experiments with more resources will explore its relation with resource
homogeneity and its statistical characterization.

Note that in the context of our experiments, early binding would still be
desirable for applications with a duration of $T_x$ long enough to make the
worse case scenario of $T_w$ negligible. In this case, applications with early
binding would have better TTC than those with late binding because of the
single pilot's larger size and therefore the greater level of concurrent
execution it would support for the application tasks. Both space and time
efficiency would be maintained as all the pilot cores would be utilized and no
walltime would be used on the target resource beyond the duration of $T_x$.

The analysis points to three main results: First, the use of execution
strategies enables the quantitative comparison among alternative couplings
between application and resources both in terms of scale and TTC (compare
Figure~\ref{fig:ttc_comparison_vertical_TTC} (a) and (b)). Usually, users
couple application and resources on the basis of conventions, via trial and
error, or in the only way allowed by the middleware they use. This limits the
scalability of applications and the effective utilization of multiple and
dynamic resources.

Second, within the given experimental parameters, the normalization of the
notoriously unpredictable queuing time~\cite{tsafrir2007backfilling} on HPC
resources is both measured and shown to depend on distributing the execution of
tasks on multiple pilots instantiated across at least three resources. This
improvement is also shown to be \B{independent} of the number of tasks, the
distribution of their durations, or the DCI on which pilots are instantiated.

Third, the experiments confirm that the AIMES middleware implementing the
abstractions defined in \S\ref{sec:abstractions} can acquire and integrate
information about the application and resource layer, and can use it to execute
application over multiple resources and DCI at scale. While other middleware
enable similar capabilities, AIMES middleware is lightweight, executed on the
users' systems, and with no requirements on the software stack installed on
each target resource.

Together, these results give users and middleware developers the opportunity to
base execution decisions on measurable performance differences of the coupling
between application and multiple dynamic resources.

\section{Conclusions}
\label{sec:conclusions}

This paper offers three main contributions to the issue of coupling distributed
application to multiple dynamic resources: (1) abstractions to represent
application's tasks, resource availability and capabilities, and execution
process; (2) the implementation of these abstractions in middleware designed to
work as a virtual laboratory for distributed application experiments; (3)
experimental comparison of execution strategies and their performance.

Together, these contributions have potential for far reaching practical and
conceptual impact. In 2005-9, an attempt was made to use multiple TeraGrid
resources to execute O(1000) parallel and concurrent
tasks~\cite{martin2009determination}. The attempt was ultimately unachievable
due to the complexity and challenges of scale. In contrast, we routinely
executed O(1000) parallel and concurrent tasks on multiple resources thanks to
more scalable infrastructure and to the effectiveness of the abstractions
underlying our implementation. Unlike other middleware, this is achieved
without requiring the installation of any software on the resources and in
compliance with their policies that don't permit services to be run on the head
nodes.

The AIMES middleware is an incubator for technologies to contribute to the HPDC
community. The interoperability layer of our middleware abstracts the
properties of diverse resources (Beowulf and Cray clusters, HTCondor pools,
Unix workstations) and it has been adopted by the LHC ATLAS experiment to
execute a mix of job sizes on DOE supercomputers~\cite{klimentov2015next}.
Analogously, RADICAL-Pilot is being used as a stand-alone pilot framework for
several project spanning molecular, bioinformatic, and earth sciences.

The AIMES middleware is also a virtual laboratory. We are extending the current
experiments and starting new ones. We have added support for distinct DCI
worldwide including OSG, XSEDE, NERSC, LRZ, UKNSS, and
FutureGrid/FutureSystems. We are extending the experiments presented in this
paper to up to 17 resources and generalizing to investigate different metrics
including throughput. We are also adding network information to the resource
bundle to experiment with execution strategies for data-intensive applications,
and we started to experiment with distributed applications comprised of
non-uniform task sizes. This will let us better understand and mimic current
DCI workloads.

In addition to exploring and measuring the dynamism of resources, we are also
planning to investigate the dynamism of distributed applications. To this end,
we have integrated Swift~\cite{wilde2011swift} with the middleware discussed in
this paper. We have begun experimenting with a greater range of applications
and with ways to decompose Swift workflows to adapt to resource availability
and capabilities. Ultimately, we will also study dynamic execution where
application strategies change during execution to maintain the coupling between
dynamic workloads and dynamic resources.

The coupling and execution of workflows with dynamic task provisioning and data
dependences will require execution strategies with large number of decisions.
For example, decisions will concern heterogeneity of the size and duration of
pilots, the evenness of their distribution across resources, the number and
preemption of their instantiation on each resource, replication of data, and
data/compute affinity. All these decisions will have to be evaluated not only
for TTC but also for other metrics involving execution reliability, resource
availability, allocation consumption, and energy efficiency.


\section*{Contributions and Acknowledgments}

MT, SJ, DSK, JW planned this paper. MT wrote it with contributions from SJ,
DSK, FL and JW. MT designed, executed, and analyzed the experiments with help
from SJ and AM. AM designed and implemented RADICAL Pilot, MT the execution
manager. DSK, MW, and ZZ designed application skeletons and ZZ implemented
them. FL and JW designed resource bundles and FL implemented them. This work is
funded by DOE ASCR, grants DE-FG02-12ER26115, DE-SC0008617, DE-SC0008651, and
NSF CAREER ACI-1253644 (SJ). This work used computing resources provided by
XRAC award TG-MCB090174 and NERSC. We thank Mark Santcroos and Yadu Nand Babuji
for useful contributions to RADICAL Pilot and skeletons respectively.  Work by
Katz was supported by NSF.  Any opinion, finding, and conclusions or
recommendations expressed in this material are those of the author(s) and do
not necessarily reflect the views of the NSF.

\def\IEEEbibitemsep{0pt plus .5pt}
\bibliographystyle{IEEEtran}
\bibliography{aimes}

\begin{thebibliography}{10}
\providecommand{\url}[1]{#1}
\csname url@samestyle\endcsname
\providecommand{\newblock}{\relax}
\providecommand{\bibinfo}[2]{#2}
\providecommand{\BIBentrySTDinterwordspacing}{\spaceskip=0pt\relax}
\providecommand{\BIBentryALTinterwordstretchfactor}{4}
\providecommand{\BIBentryALTinterwordspacing}{\spaceskip=\fontdimen2\font plus
\BIBentryALTinterwordstretchfactor\fontdimen3\font minus
  \fontdimen4\font\relax}
\providecommand{\BIBforeignlanguage}[2]{{%
\expandafter\ifx\csname l@#1\endcsname\relax
\typeout{** WARNING: IEEEtran.bst: No hyphenation pattern has been}%
\typeout{** loaded for the language `#1'. Using the pattern for}%
\typeout{** the default language instead.}%
\else
\language=\csname l@#1\endcsname
\fi
#2}}
\providecommand{\BIBdecl}{\relax}
\BIBdecl

\bibitem{lsst-url}
Large Synoptic Survey Telescope, \url{http://www.lsst.org/lsst/}.

\bibitem{ska-url}
The Square Kilometer Array, \url{https://www.skatelescope.org/}.

\bibitem{lhc-url}
{LHC}, \url{http://cern.ch/topics/large-hadron-collider}.

\bibitem{hey2009fourth}
A.~J. Hey, S.~Tansley, K.~M. Tolle \emph{et~al.}, \emph{The fourth paradigm:
  data-intensive scientific discovery}.\hskip 1em plus 0.5em minus 0.4em\relax
  MS Research, 2009.

\bibitem{hpdc_papers_url}
Best Papers of {HPDC}, \url{http://www.hpdc.org/best.php/}.

\bibitem{katz2011cyberinfrastructure}
D.~S. Katz, D.~Hart, C.~Jordan, A.~Majumdar, J.-P. Navarro, W.~Smith, J.~Towns,
  V.~Welch, and N.~Wilkins-Diehr, ``Cyberinfrastructure usage modalities on the
  teragrid,'' in \emph{Proc. IEEE Int. Parallel \& Distributed Processing
  Symp.}, 2011, pp. 932--939.

\bibitem{foster1996software}
I.~Foster, J.~Geisler, B.~Nickless, W.~Smith, and S.~Tuecke, ``Software
  infrastructure for the {I-WAY} high performance distributed computing
  experiment,'' in \emph{Proc. 5th IEEE Symp. on HPDC}, 1996, pp. 562--571.

\bibitem{grimshaw1997legion}
A.~S. Grimshaw, W.~A. Wulf \emph{et~al.}, ``The legion vision of a worldwide
  virtual computer,'' \emph{Comm. of the ACM}, vol.~40, no.~1, pp. 39--45,
  1997.

\bibitem{foster2001anatomy}
I.~Foster, C.~Kesselman, and S.~Tuecke, ``The anatomy of the grid: Enabling
  scalable virtual organizations,'' \emph{Int. journal of HPC applications},
  vol.~15, no.~3, pp. 200--222, 2001.

\bibitem{thain2005distributed}
D.~Thain, T.~Tannenbaum, and M.~Livny, ``Distributed computing in practice: The
  condor experience,'' \emph{CCPE}, vol.~17, no. 2-4, pp. 323--356, 2005.

\bibitem{allen2001supporting}
G.~Allen, I.~Foster, N.~T. Karonis, M.~Ripeanu, E.~Seidel, and B.~Toonen,
  ``Supporting efficient execution in heterogeneous distributed computing
  environments with {Cactus} and {Globus},'' in \emph{Proc. ACM/IEEE Conf. on
  Supercomputing}, 2001, p.~52.

\bibitem{gardner2006parallel}
M.~K. Gardner, W.-c. Feng, J.~Archuleta, H.~Lin, and X.~Ma, ``Parallel genomic
  sequence-searching on an ad-hoc grid: experiences, lessons learned, and
  implications,'' in \emph{Proc. ACM/IEEE SC Conf.}, 2006, pp. 22--22.

\bibitem{Jha2013distributed}
S.~Jha, M.~Cole, D.~S. Katz, M.~Parashar, O.~Rana, and J.~Weissman,
  ``Distributed computing practice for large-scale science and engineering
  applications,'' \emph{Concurrency and Computation: Practice and Experience},
  vol.~25, no.~11, pp. 1559--1585, 2013.

\bibitem{bokhari1981shortest}
S.~H. Bokhari, ``A shortest tree algorithm for optimal assignments across space
  and time in a distributed processor system,'' \emph{IEEE Trans. on Software
  Engineering}, vol. SE-7, no.~6, pp. 583--589, 1981.

\bibitem{fernandez1989allocating}
D.~Fern{\'a}ndez-Baca, ``Allocating modules to processors in a distributed
  system,'' \emph{IEEE Trans. on Software Engineering}, vol.~15, no.~11, pp.
  1427--1436, 1989.

\bibitem{chen2012partitioning}
W.~Chen and E.~Deelman, ``Partitioning and scheduling workflows across multiple
  sites with storage constraints,'' in \emph{Parallel Processing and Applied
  Mathematics}, 2012, pp. 11--20.

\bibitem{malawski2015algorithms}
M.~Malawski, G.~Juve, E.~Deelman, and J.~Nabrzyski, ``Algorithms for cost-and
  deadline-constrained provisioning for scientific workflow ensembles in iaas
  clouds,'' \emph{Future Generation Computer Systems}, vol.~48, pp. 1--18,
  2015.

\bibitem{taylor2006}
I.~J. Taylor, E.~Deelman, D.~B. Gannon, and M.~Shields, \emph{Workflows for
  e-Science: Scientific Workflows for Grids}.\hskip 1em plus 0.5em minus
  0.4em\relax Secaucus, NJ, USA: Springer-Verlag New York, Inc., 2006.

\bibitem{foster1990parallel}
I.~T. Foster and R.~L. Stevens, ``Parallel programming with algorithmic
  motifs,'' in \emph{Proc. Int. Conf. on Parallel Processing}, vol. 2:
  Software, 1990, pp. 26--34.

\bibitem{meyerwgl}
L.~Meyer, M.~Mattoso, M.~Wilde, and I.~Foster, ``Wgl -- a workflow generator
  language and utility,'' University of Chicago, Tech. Rep., 2013.

\bibitem{cardosa2008resource}
M.~Cardosa and A.~Chandra, ``Resource bundles: Using aggregation for
  statistical wide-area resource discovery and allocation,'' in \emph{Proc.
  28th Int. Conf. on Distributed Computing Systems}, 2008, pp. 760--768.

\bibitem{liu2003constraint}
C.~Liu and I.~Foster, ``A constraint language approach to grid resource
  selection,'' in \emph{Proceeding of HPDC}, 2003.

\bibitem{oppenheimer2005design}
D.~Oppenheimer, J.~Albrecht, D.~Patterson, and A.~Vahdat, ``Design and
  implementation tradeoffs for wide-area resource discovery,'' in \emph{Proc.
  14th IEEE Int. Symp. on HPDC}, 2005, pp. 113--124.

\bibitem{nurmi2008qbets}
D.~Nurmi, J.~Brevik, and R.~Wolski, ``{QBETS}: Queue bounds estimation from
  time series,'' in \emph{Job Scheduling Strategies for Parallel Processing},
  2008.

\bibitem{tsafrir2007backfilling}
D.~Tsafrir, Y.~Etsion, and D.~G. Feitelson, ``Backfilling using
  system-generated predictions rather than user runtime estimates,'' \emph{IEEE
  Trans. on Par. and Distr. Systems}, 2007.

\bibitem{raicu2008many}
I.~Raicu, I.~T. Foster, and Y.~Zhao, ``Many-task computing for grids and
  supercomputers,'' in \emph{Proc. Wksp on Many-Task Computing on Grids and
  Supercomputers}, 2008, pp. 1--11.

\bibitem{zhang2014using}
Z.~Zhang and D.~S. Katz, ``Using application skeletons to improve escience
  infrastructure,'' in \emph{Proc. 10th IEEE Int. Conf. on e-Science}, vol.~1,
  2014, pp. 111--118.

\bibitem{skeleton-FGCS}
D.~S. Katz, A.~Merzky, Z.~Zhang, and S.~Jha, ``Application skeletons:
  Construction and use in {eScience},'' \emph{Future Generation Computer
  Systems}, 2015.

\bibitem{skeleton-code-v1.2}
D.~S. Katz, A.~Merzky, M.~Turilli, M.~Wilde, and Z.~Zhang, ``Application
  skeleton v1.2,'' \emph{GitHub}, Jan 2015.

\bibitem{deelman2005pegasus}
E.~Deelman, G.~Singh, M.-H. Su, J.~Blythe, Y.~Gil, C.~Kesselman, G.~Mehta,
  K.~Vahl, G.~B. Berriman, J.~Good, A.~Laity, J.~C. Jacob, and D.~S. Katz,
  ``Pegasus: A framework for mapping complex scientific workflows onto
  distributed systems,'' \emph{Scientific Programming}, vol.~13, no.~3, pp.
  219--237, 2005.

\bibitem{wilde2011swift}
M.~Wilde, M.~Hategan, J.~M. Wozniak, B.~Clifford, D.~S. Katz, and I.~Foster,
  ``Swift: {A} language for distributed parallel scripting,'' \emph{Parallel
  Computing}, pp. 633--652, Sept. 2011.

\bibitem{jacob2009montage}
J.~C. Jacob, D.~S. Katz, G.~B. Berriman, J.~C. Good, A.~Laity, E.~Deelman,
  C.~Kesselman, G.~Singh, M.-H. Su, T.~Prince \emph{et~al.}, ``Montage: a grid
  portal and software toolkit for science-grade astronomical image
  mosaicking,'' \emph{Int. Journal of Computational Science and Engineering},
  vol.~4, no.~2, pp. 73--87, 2009.

\bibitem{mathog2003parallel}
D.~R. Mathog, ``Parallel {BLAST} on split databases,'' \emph{Bioinformatics},
  vol.~19, no.~14, pp. 1865--1866, 2003.

\bibitem{maechling2007scec}
P.~Maechling \emph{et~al.}, ``{SCEC} {CyberShake} workflows—automating
  probabilistic seismic hazard analysis calculations,'' in \emph{Workflows for
  e-Science}, 2007, pp. 143--163.

\bibitem{simmhan2010comparison}
Y.~Simmhan and L.~Ramakrishnan, ``Comparison of resource platform selection
  approaches for scientific workflows,'' in \emph{Proc. 19th ACM Int. Symp. on
  HPDC}, 2010, pp. 445--450.

\bibitem{downey1997predicting}
A.~B. Downey, ``Predicting queue times on space-sharing parallel computers,''
  in \emph{Proc. 11th Int. Parallel Processing Symp.}, 1997.

\bibitem{vazhkudai2001predicting}
S.~Vazhkudai, J.~M. Schopf, and I.~Foster, ``Predicting the performance of wide
  area data transfers,'' in \emph{Proc. 16th IEEE Int. Parallel and Distributed
  Processing Symp.}, 2001.

\bibitem{kosar2004stork}
T.~Kosar and M.~Livny, ``Stork: Making data placement a first class citizen in
  the grid,'' in \emph{Proc. 24th IEEE Int. Conf. on Distributed Computing
  Systems}, 2004.

\bibitem{raghavan2014tiera}
A.~Raghavan, A.~Chandra, and J.~Weissman, ``Tiera: towards flexible
  multi-tiered cloud storage instances,'' in \emph{Proc. 15th Int. Middleware
  Conf.}, 2014, pp. 1--12.

\bibitem{atlas2008atlas}
{ATLAS collaboration}, G.~Aad \emph{et~al.}, ``The {ATLAS} experiment at the
  {CERN} large hadron collider,'' \emph{JINST}, vol.~3, 2008.

\bibitem{maeno2014evolution}
T.~Maeno, K.~De, A.~Klimentov, P.~Nilsson, D.~Oleynik, S.~Panitkin,
  A.~Petrosyan, J.~Schovancova, A.~Vaniachine, and T.~Wenaus, ``Evolution of
  the {ATLAS PanDA} workload management system for exascale computational
  science,'' \emph{J. Phys.: Conf. Ser.}, vol. 513, p. 032062, 2014.

\bibitem{pordes2007open}
R.~Pordes \emph{et~al.}, ``The open science grid,'' \emph{J. Phys.: Conf.
  Ser.}, vol.~78, no.~1, p. 012057, 2007.

\bibitem{katz2012survey}
D.~S. Katz, S.~Jha, M.~Parashar, O.~Rana, and J.~Weissman, ``Survey and
  analysis of production distributed computing infrastructures,'' \emph{arXiv
  preprint arXiv:1208.2649}, 2012.

\bibitem{hategan2011coasters}
M.~Hategan, J.~Wozniak, and K.~Maheshwari, ``Coasters: uniform resource
  provisioning and access for clouds and grids,'' in \emph{Proc. 4th IEEE Int.
  Conf. on Utility and Cloud Computing}, 2011, pp. 114--121.

\bibitem{jain2015fireworks}
A.~Jain, S.~P. Ong, W.~Chen, B.~Medasani, X.~Qu, M.~Kocher, M.~Brafman,
  G.~Petretto, G.-M. Rignanese, G.~Hautier \emph{et~al.}, ``{FireWorks}: a
  dynamic workflow system designed for high-throughput applications,''
  \emph{Concurrency and Computation: Practice and Experience}, 2015.

\bibitem{review_radicalpilot_2015}
A.~Merzky, M.~Santcroos, M.~Turilli, and S.~Jha, ``{RADICAL-Pilot: Scalable
  Execution of Heterogeneous and Dynamic Workloads on Supercomputers},'' 2015,
  (under review) \url{http://arxiv.org/abs/1512.08194}.

\bibitem{saga-x}
A.~Merzky, O.~Weidner, and S.~Jha, ``{SAGA}: A standardized access layer to
  heterogeneous distributed computing infrastructure,'' \emph{Software-X}, vol.
  1-2, pp. 1--3, 2015.

\bibitem{goodale2008simple}
T.~Goodale, S.~Jha, H.~Kaiser, T.~Kielmann, P.~Kleijer, A.~Merzky, J.~Shalf,
  and C.~Smith, ``A simple {API} for grid applications ({SAGA}),'' Open Grid
  Forum, {OGF} Recommendation, {GFD.90}, 2007.

\bibitem{piloterview-2015}
\BIBentryALTinterwordspacing
M.~Turilli, M.~Santcroos, and S.~Jha, ``{A Comprehensive Perspective on
  Pilot-Abstraction},'' 2015. [Online]. Available:
  \url{http://arxiv.org/abs/1508.04180}
\BIBentrySTDinterwordspacing

\bibitem{bfe-jctc-2014}
D.~W. Wright, B.~A. Hall, O.~A. Kenway, S.~Jha, and P.~V. Coveney, ``Computing
  clinically relevant binding free energies of {HIV-1} protease inhibitors,''
  \emph{Chemical Theory and Computation}, vol.~10, no.~3, pp. 1228--1241, 2014.

\bibitem{furlani2013using}
T.~R. Furlani, B.~L. Schneider, M.~D. Jones, J.~Towns, D.~L. Hart, S.~M. Gallo,
  R.~L. DeLeon, C.-D. Lu, A.~Ghadersohi, R.~J. Gentner \emph{et~al.}, ``Using
  xdmod to facilitate xsede operations, planning and analysis,'' in \emph{Proc.
  of XSEDE13: Gateway to Discovery}.\hskip 1em plus 0.5em minus 0.4em\relax
  ACM, 2013, p.~46.

\bibitem{martin2009determination}
H.~S. Martin, S.~Jha, S.~Howorka, and P.~V. Coveney, ``Determination of free
  energy profiles for the translocation of polynucleotides through
  $\alpha$-hemolysin nanopores using non-equilibrium molecular dynamics
  simulations,'' \emph{J. Chem. Theory Comput.}, vol.~5, no.~8, pp. 2135--2148,
  2009.

\bibitem{klimentov2015next}
A.~Klimentov \emph{et~al.}, ``Next generation workload management system for
  big data on heterogeneous distributed computing,'' \emph{J. Phys.: Conf.
  Ser.}, vol. 608, p. 012040, 2015.

\end{thebibliography}

\end{document}